\documentclass[11pt,a4paper]{article}
\usepackage{jcappub}
\usepackage{color,amsmath,natbib,latexsym,verbatim,hyperref}

\definecolor{MyDarkBlue}{rgb}{0,0.08,0.45}

\title{Searches for Dark Matter annihilation signatures in the Segue~1
  satellite galaxy with the MAGIC-I telescope}  

\collaboration{MAGIC Collaboration}

\author[1]{J.~Aleksi\'c}
\author[2]{E.~A.~Alvarez}
\author[3]{L.~A.~Antonelli}
\author[4]{P.~Antoranz}
\author[2]{M.~Asensio}
\author[5]{M.~Backes}
\author[2]{J.~A.~Barrio}
\author[6]{D.~Bastieri}
\author[7,8]{J.~Becerra Gonz\'alez}
\author[9]{W.~Bednarek}
% ...

\affiliation[1]{IFAE, Edifici Cn., Campus UAB, E-08193 Bellaterra, Spain}
\affiliation[2]{Universidad Complutense, E-28040 Madrid, Spain}
\affiliation[3]{INAF National Institute for Astrophysics, I-00136 Rome, Italy}
\affiliation[4]{Universit\`a  di Siena, and INFN Pisa, I-53100 Siena, Italy}
\affiliation[5]{Technische Universit\"at Dortmund, D-44221 Dortmund, Germany}
\affiliation[6]{Universit\`a di Padova and INFN, I-35131 Padova, Italy}
\affiliation[7]{Inst. de Astrof\'{\i}sica de Canarias, E-38200 La Laguna, Tenerife, Spain}
\affiliation[8]{Depto. de Astrof\'{\i}sica, Universidad de La Laguna, E-38206 La Laguna, Spain}
\affiliation[9]{University of \L\'od\'z, PL-90236 Lodz, Poland}
% ...

\emailAdd{michele.doro@uab.cat}
\emailAdd{fornasam@gmail.com}
\emailAdd{slombard@pd.infn.it}
\emailAdd{nieto@gae.ucm.es}

\abstract{
We report the results of the observation of the nearby satellite
galaxy Segue~1 performed by the MAGIC-I ground-based gamma-ray
telescope between November 2008 and March 2009 for a total of
43.2~hours. No significant gamma-ray emission was found above the
background. Differential upper limits on the gamma-ray flux are
derived assuming various power-law slopes for the possible emission
spectrum. Integral upper limits are also calculated for several
power-law spectra and for different energy thresholds. The values are
of the order of $10^{-11}$ ph cm$^{-2}$ s$^{-1}$ above 100~GeV and
$10^{-12}$ ph cm$^{-2}$ s$^{-1}$ above 200~GeV. Segue~1 is currently
considered one of the most interesting targets for indirect dark
matter searches. In these terms, the upper limits have been also
interpreted in the context of annihilating dark matter particles. For
such purpose, we performed a 
grid scan over a reasonable portion of the parameter space for the
minimal SuperGravity model and computed the flux upper limit for each
point separately, taking fully into account the peculiar 
spectral features of each model.  We found that in order to match the
experimental upper limits with the model predictions, a minimum flux
boost of $10^{3}$ is required, and that the upper limits are quite
dependent on the shape of the gamma-ray energy spectrum predicted by
each specific model.  
Finally we compared the upper limits with the predictions of some dark
matter models able to explain the PAMELA rise in the positron ratio,
finding that Segue~1 data are in tension with the dark matter
explanation of the PAMELA spectrum in the case of a dark matter
candidate annihilating into $\tau^+\tau^-$. A complete exclusion however is
not possible due to the uncertainties in the Segue 1 astrophysical
factor. 
}


\arxivnumber{1103.0477}

\notoc

\begin{document}
\maketitle

\section{Introduction}
A major open question for modern physics is the nature of dark matter
(DM): strong experimental evidences suggest the presence of this elusive 
component in the energy budget of the Universe 
\citep[see, e.g.,][]{Komatsu:2010a,Clowe:2006a}, without, however, 
being able to provide conclusive results about its nature. One of the most 
popular scenarios is that of weakly interacting massive particles (WIMPs), 
that includes a large class of non-baryonic DM candidates with a mass
typically between few tens of GeV and few TeV and an annihilation
cross section of the order of weak interactions
\citep{Bertone:2005a,ParticleDarkMatter:2010a}.
%, Bergstrom:2000, Bergstrom:2009ib}.
Natural WIMP candidates are found, e.g., in SUperSYmmetric (SUSY) extensions 
of the Standard Model (SM)~\citep{Jungman:1996a,Martin:1998a}, and
in theories with Universal Extra-Dimensions
\citep{Appelquist:2001a,Servant:2003a,Cheng:2002a}.
Recently, a number of experimental results have
appeared, which may be interpreted in terms of DM 
\citep{Colafrancesco:2010a}: in particular, the measurement of the
positron fraction in cosmic-rays in the $10-100$~GeV range by the
PAMELA satellite \citep{Adriani:2009a}, and the energy spectra of electrons 
and positrons above $300$~GeV measured by Fermi/LAT, H.E.S.S., and ATIC 
\citep{Abdo:2009a,Aharonian:2009a,Chang:2008a}.
Lately, also the search for DM with the use of underground recoil
experiments has produced interesting results 
\citep{Bernabei:2008yi,Ahmed:2009a,Aalseth:2010a,Aprile:2010a}.
In this paper we will discuss the possibility of detecting signatures of 
DM annihilation in the very high energy band (i.e. above 100~GeV) of the 
electromagnetic spectrum using as particularly convenient target the 
Segue~1 dwarf Spheroidal galaxy (dSph). 

Assuming the $\Lambda$CDM cosmological model, DM structures form by
hierachical collapse of small overdensities and are supposed to extend
in mass down to the scale of the Earth or even below
\citep{Green:2003a,Diemand:2008a,Springel:2008b}. DM structures
may also host smaller satellite structures and it has been proposed
that the dSphs in the Milky Way (MW) may have formed within some of
these subhalos hosted in the larger MW DM halo
\citep{Klypin:1999uc,Moore:1999wf,Kravtsov:2004cm,Strigari:2007ma}.  
So far, around two dozen dSphs have been identified in the MW. They
represent excellent targets for indirect DM searches due to their
very large mass-to-light ratios, low baryonic content (which
disfavours gamma-ray emission from conventional astrophysical sources)
and their location often at high galactic latitudes, where
contamination by Galactic background is subdominant
\citep{Gilmore:2007a,Simon:2007a,Strigari:2007a,SanchezConde:2007a}.

To this class belongs Segue~1, a satellite galaxy discovered
by Belokurov et al.~\citep{Belokurov:2007a} in the Sloan Extension for Galactic
Understanding and Exploration \citep[SEGUE,][]{Yanny:2009a}. It is
located at a distance of $23\pm2$~kpc from the Sun ($28$~kpc from the Galactic
Center) at (RA, Dec) = ($10.12^h$, $16.08^\circ$). Despite the fact that its
nature has been debated after discovery
\citep{Belokurov:2007a,NiedersteOstholt:2009a,Geha:2009a,XiangGruess:2009a},
it has now been interpreted more clearly as an ultrafaint MW satellite
galaxy in Ref.~\citep{Simon:2010a} through the identification of several new
member stars (66 instead of the previous 24). This interpretation has 
also been confirmed in Ref.~\citep{Martinez:2010a} using a Bayesian analysis of
star membership probabilities. 
Segue~1 was highlighted as a good target for indirect DM detection in Refs.
\citep{Essig:2009a, Martinez:2009jh}.  We will show
later on in section~\ref{sec:model} that, following our calculations, Segue~1
results to be the most DM dominated dwarf in our galaxy, known so far. 
The Fermi/LAT
\citep{Atwood:2009a} has observed Segue~1 in its survey observation 
mode: results from the first 3 months of data have been presented in
Refs.~\citep{Farnier:2009a}, while results from the 
first 9 months of data have been presented in
Refs.~\citep{Murgia:2009, Jeltema:2009}. No gamma-ray signal was detected. Implications
for DM models have been considered in Refs.~\citep{Scott:2010a, Essig:2010a}.
Segue~1 was also observed
by the X-ray telescope onboard the SWIFT satellite. Data from this short 
observation ($\sim5$~ks) have been analyzed \citep{Mirabal:2010a} 
where no detection is reported either.

We present here the results of the observation
of the Segue~1 satellite performed by the Major Atmospheric Gamma-ray 
Imaging Cherenkov (MAGIC) telescope. The observation took place 
between November 2008 and March 2009, for a total of $43.2$~hours ($29.4$~hours in
good observational conditions after quality cuts). No significant detection 
above the background was found for energies above 100 GeV. In this paper 
we describe the data analysis and determine upper limits (ULs) for the
gamma-ray emission. We extensively study the dependence of the ULs on 
a set of parameters like the expected energy spectrum of the source or 
on experimental features like the energy threshold and the angular 
resolution. 
The MAGIC-I observation is then used to put constraints on some models
of DM. We focus on the case of annihilating SUSY DM candidates and 
compute a grid scan over a reasonable range of the SUSY parameters.
Point per point we use the predicted energy spectrum to compute
specific ULs, showing that the dependence on the energy spectrum is even
more prominent than in the case of power-law spectra.
ULs are also given for DM models that fit the PAMELA and Fermi/LAT
cosmic-ray spectra.

The paper is organized as follows: in section \ref{sec:magic} we
describe the observation of Segue~1 with MAGIC-I and discuss the analysis.
In section \ref{sec:results} the calculation of the ULs on the differential 
and integral flux is outlined and results are reported in the case of
power-law spectra. 
%Preliminary estimation for the 
%MAGIC stereo system are also considered.
We focus on a possible DM interpretation in section \ref{sec:model}. 
Conclusions are drawn in section \ref{sec:summary} together with a brief 
summary.

\section{Observation of Segue~1 with the MAGIC-I telescope}
\label{sec:magic}
The MAGIC experiment for ground-based gamma-ray astronomy consists of
a system of two telescopes operating in stereoscopic mode since fall
2009 at the Canary Island of La Palma ($28.8^{\circ}$~N,
$17.8^{\circ}$~W, 2200~m~a.s.l.). However, the observation of Segue~1
was carried out when only the first 
telescope was operating. The data were taken during dark nights
between November 2008 and March 2009, for a total exposure time of
$43.2$ hours. The source was surveyed at zenith angles between
$12.7^\circ$ and $33.9^\circ$, which guarantees a low energy
threshold, in false-source tracking (\emph{wobble})
mode~\citep{Fomin:1994a}, in which the pointing direction alternates
every 20 minutes between two positions, offset by $\pm0.4^{\circ}$ in
RA from the source.  

The data analysis was performed using the standard MAGIC-I analysis
and reconstruction software. After the calibration and image
cleaning~\citep{Albert:2008c,Albert:2008b}, the hadronic background
rejection is achieved through a multivariate method called Random
Forest (RF)~\citep{Breiman:2001a,Albert:2008b}. The algorithm uses the
image Hillas parameters~\citep{Hillas:1985a} and timing variables to
compute a gamma-ray/hadron discriminator called \emph{hadronness} by 
comparison with Monte Carlo (MC) gamma-ray simulations. Hadronness ranges 
from $0$ (for showers confidently identified as initiated by gamma-rays) 
to $1$ (for those clearly showing the features of a hadronic cosmic-ray 
initiated shower). The RF method is also used to estimate the energy of the
showers. The detector MC simulation was tuned to fit the
actual telescope performance at the time of the observations.  
After standard quality cuts in trigger rate, atmospheric condition,
and a check of the distribution of basic image parameters, around 30\%
of the Segue~1 data were rejected, resulting in $29.4$~hours of data. The 
same quality cuts were applied to a sample of Crab Nebula data, taken 
in similar experimental conditions, to cross-check the RF routines and the 
analysis cuts. 

With respect to the standard MAGIC-I analysis, an additional cut was
introduced to account for the presence of the star $\eta-$Leonis, with
apparent magnitude $V=3.5$, $B-V=0.02$ and
(RA, Dec) = ($10.12^h$, $16.76^\circ$), located at $0.68^\circ$ 
from the position of Segue~1. The light of the star entered the trigger 
region of the MAGIC-I camera and, as a consequence, produced local 
inhomogeneities, which are not treated  automatically  by the standard analysis 
at energies below 200~GeV. Therefore, all events below this energy and 
spatially coincident with an optimized region around the position of the 
star in the camera were discarded. With such additional \emph{star-cuts},
around $20$\% of the events were rejected,  but the local
inhomogeneities caused by the star in the trigger area were
efficiently washed out. To check the
effect of the cuts in the effective area 
and eventually in the sensitivity,  the same star-cuts were
applied to the Crab Nebula data, and to the MC~gamma-ray events for 
consistency. The final energy threshold was at 100~GeV and the star-cuts 
had the main effect of slightly reducing the effective area at lower
energies. 

%, as shown in figure~\ref{fig:effective_area}. 
%\begin{figure}[t]
%\centering
%\includegraphics[width=0.95\linewidth]{./Aeff_stardistcuts.eps}
%\caption{\label{fig:effective_area} Effective collection area above
%  100 GeV as a function of the true MC energy, before and after the
%  analysis cuts (blue and green points) and including also the
%  star-cuts (red points, see text for details).} 
%\end{figure}

\begin{figure}[t]
\centering
\includegraphics[width=0.95\linewidth]{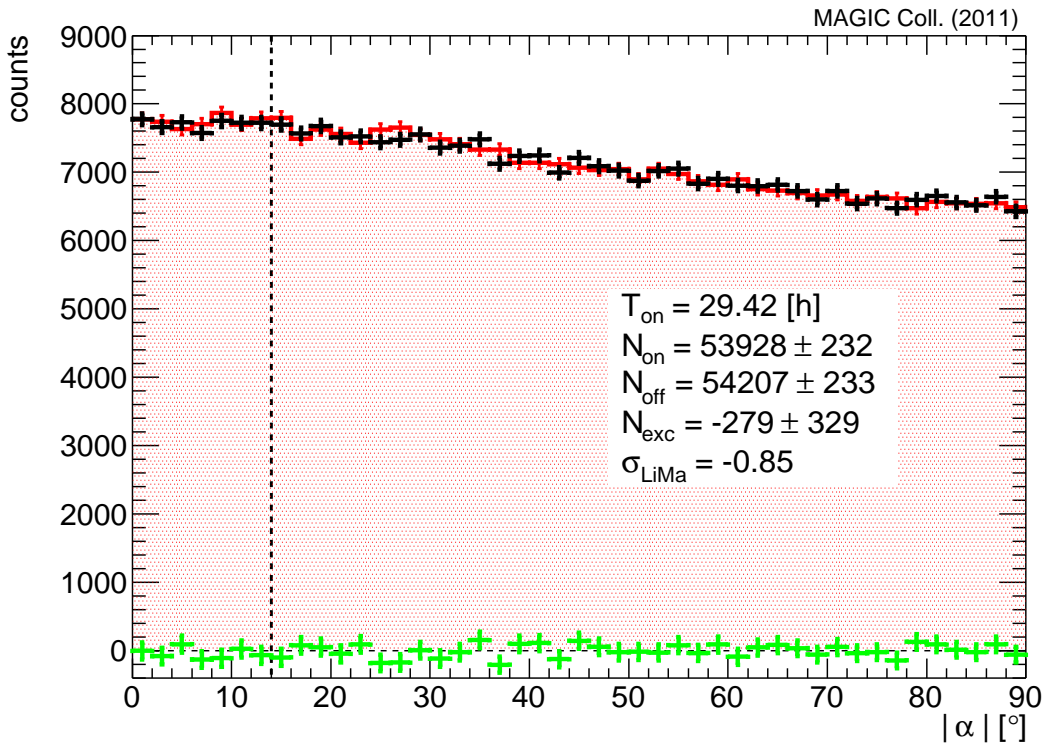}
\caption{$|\alpha|-$plot from $29.4$ hours of observation of Segue~1
  above 100~GeV with the MAGIC-I telescope. The vertical dashed line 
  represents the fiducial region for the static cut of
  $|\alpha|<14^{\circ}$, where the signal is  
  expected. Red points represent the signal (ON sample), black points
  the background (OFF sample), and green points their difference.} 
\label{fig:alpha_plot}
\end{figure}

The number of candidates for gamma-ray events from the direction of the 
source was estimated using the distribution of $|\alpha|$
angles\footnote{The $|\alpha|$ parameter is defined as the angle
  between the major axis of an image and the direction from the image
  center of gravity to a reference point in the field-of-view, which is
  the nominal source position for the ON data, and a control
  background point for the OFF data~\citep{Hillas:1985a}}, which is
related to the orientation of the showers. The best set of cuts for the hadronness 
and $|\alpha|$ parameters were optimized on the Crab Nebula data sample, 
separately in logarithmic energy bins  (dynamic cuts),  and under the 
assumption of a point-like emission. In section~\ref{sec:model} we will 
further discuss this issue. In figure~\ref{fig:alpha_plot}, the $|\alpha|$-plot 
above 100~GeV is shown.  For this plot, the  number of excess events was computed  
in a fixed fiducial signal region  (static cut)  with
$|\alpha|<14^{\circ}$\footnote{ The static $|\alpha|$-cut at $14^\circ$
  is used for display purpose. In the computation of the flux upper
  limits, the $|\alpha|$-cuts are instead optimized more suitably in a
  dynamic way in each energy bin. In the first two bins, were most of
  the events are concentrated, the optimized $|\alpha|$-cut is
  $14^\circ$. }  
%  , while at higher energies, this cut gets stronger because of the
%  better instrumental angular resolution. 
and is $N_{exc}$($>100$~GeV)$=-279\pm329$, corresponding to a 
significance of $-0.85\;\sigma$, computed using eq.~(17) of Li \& Ma~\citep{Li:1983a}.

%%%%%%%%%% FIGURE 1 %%%%%%%%%%%%

\begin{figure}[t]
\centering
\includegraphics[width=0.95\linewidth]{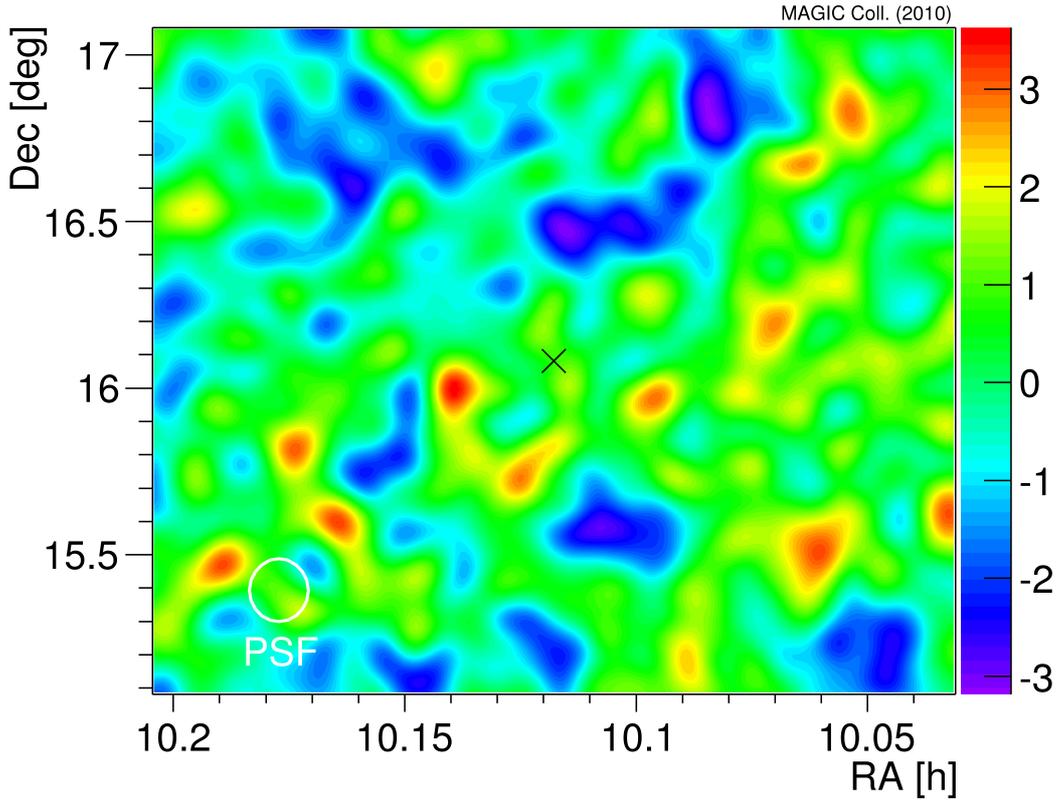}
\caption{\label{fig:skymap}Significance map for events above 200~GeV in the 
  Segue~1 sky region. The black cross marks the position of the
  central core of Segue~1
  and the telescope PSF is also shown. The significance  distribution is 
  consistent with background fluctuations.} 
\end{figure}

In figure~\ref{fig:skymap}, we show the significance map for the sky
region around Segue~1. For this plot the source independent DISP
method was used \citep{DomingoSantamaria:2005a}. The energy threshold in this
case was raised to 200~GeV to cope with the presence of
the star in the trigger region. The significance distribution is
consistent with background fluctuations. 

%%%%%%%%%% FIGURE 2 %%%%%%%%%%%%

%=======================================================================
\section{Upper Limits for Power-Law Spectra}
\label{sec:results}
In the previous section we presented the Segue~1 data collected by the
MAGIC-I telescope, showing that, above 100~GeV, results are 
consistent with no signal over the background. We derive now
differential and integral flux ULs for the gamma-ray emission from the 
source. 
%We will pay special attention to the dependence of ULs on different 
%assumptions and the robustness of the results.
In the computation of the ULs, we will distinguish
between ``true'' energy $E$ (defined for the MC gamma-ray
simulated events) and ``reconstructed'' energy $E_{rec}$, calculated by
means of the RF method (see section~\ref{sec:magic}) for the real as
well as for the MC gamma-ray data. 

% Differential upper limits
% ======================================
\subsection{Differential Upper Limits for power-law spectra}
In order to calculate the differential flux UL in an energy bin
$\Delta E_{rec}$ one can estimate the UL in the number of excess
events $N^{UL}_{exc}(\Delta E_{rec})$ following the so-called ``Rolke
method'' \citep[assuming 30\% systematic uncertainty in the signal
  efficiency and 95\%
  confidence level]{Rolke:2005a} from the $|\alpha|-$plot (obtained
after the usual analysis cuts \emph{and} only for reconstructed
energies within the bin $\Delta E_{rec}$). The average effective area
$A_{eff}(E;\Delta E_{rec})$ is calculated from the effective area for
gamma-rays of true energy $E$, after all analysis cuts, including
$E_{rec}\in\Delta E_{rec}$, and convolved with the energy spectrum. The latter, in this case, is  assumed to be a  power-law of the form
$S(E)\propto(E/E_*)^{\Gamma}$, $\Gamma$ being the spectral index and
$E_*$ the pivot energy for the particular energy bin, defined as:
\begin{equation}
%E_{*} = \frac{ \int_{0}^{\infty} E \; E^{\Gamma} \;
%    A_{eff}(E; \Delta E_{rec}) dE} {\int_{0}^{\infty}
%  E^{\Gamma}  \; A_{eff}(E; \Delta E_{rec}) dE}.
E_{*} = \frac{ \int_{0}^{\infty} E \; S(E) \;
    A_{eff}(E; \Delta E_{rec}) dE} {\int_{0}^{\infty}
  S(E) \; A_{eff}(E; \Delta E_{rec}) dE}.
\label{eq:Emean}
\end{equation}
Finally, the differential flux UL reads as follows:

\begin{equation}
%  \frac{d\Phi^{UL}}{dE}(E_*) 
%  = \frac{N^{UL}_{exc}(\Delta E_{rec})}{t_{eff}}
%  \frac{1}{\int_{0}^{\infty} A_{eff}(E; \Delta E_{rec}) \;
%       (E/E_*)^\Gamma dE}\; 
  \frac{d\Phi^{UL}}{dE}(E_*) 
  = \frac{N^{UL}_{exc}(\Delta E_{rec})}{t_{eff}}
  \frac{1}{\int_{0}^{\infty} A_{eff}(E; \Delta E_{rec}) \;
       S(E)\;dE}\; 
\label{eq:diffupperlimits}
\end{equation}
and is computed as the UL in the number of excess events divided by the effective
observation time and the average effective area weighted with the
expected gamma-ray spectrum. Thus, results depend on the
assumed source spectral shape (although this dependence gets less
significant as the binning goes finer).  
%In the second term of the equation is the average effective area
%reweighted with the gamma-ray spectrum, and at the numerator
%$S(E_*)=(E/E_*)^\Gamma=1$.  
The differential UL is measured in ph
cm$^{-2}$ s$^{-1}$ TeV$^{-1}$.

\begin{figure}[t]
\centering
\includegraphics[width=0.95\linewidth]{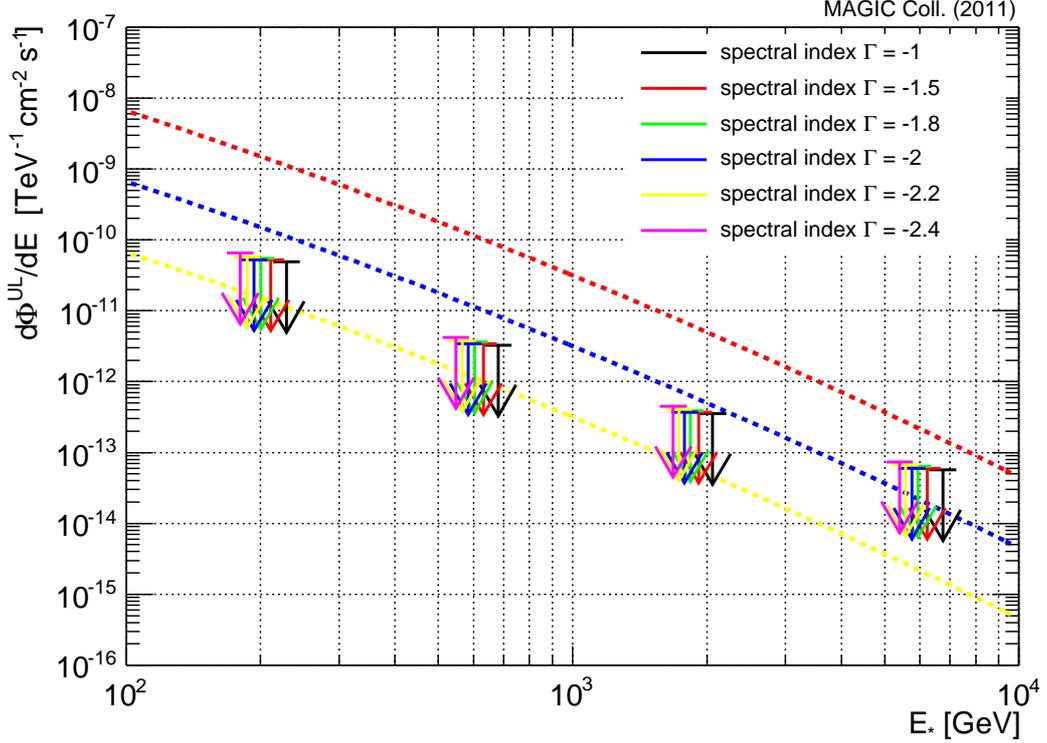}
\caption{Differential flux upper limits from Segue~1 as in 
  table~\ref{tab:uldiff}.  As reference, the Crab Nebula differential
  flux (red dashed line) \cite{Albert:2007xh} and its 10\% (blue dashed line) and 1\%
  (yellow dashed line) fractions are also drawn.}  
\label{fig:uldiff}
\end{figure}

Table~\ref{tab:uldiff} summarizes the ULs in four
reconstructed--energy logarithmic  
bins between 100~GeV and 10~TeV using eq.~(\ref{eq:diffupperlimits}) and
assuming different power-law spectra, with spectral indexes
$\Gamma=-1.0, -1.5, -1.8, -2.0, -2.2, -2.4$ respectively,  as
done in Ref.~\cite{Abdo:2010b}. Moreover, the case of $\Gamma=-1.5$ is
considered, as a reference for hadronic annihilation DM models
\cite{Fornengo:2004kj}.  The results are
also shown in  figure~\ref{fig:uldiff}.
% The differential ULs on the flux are more and more independent from
%the assumed slope as the energy bins considered grows smaller and
%smaller.  
%, if computed in 
%energy bins that are small enough, result to be quite independent on the 
%assumed slope. 

\begin{table}[!htb]
\scriptsize{%
\centering
\caption{Differential Segue~1 flux upper limits for several 
  power-law gamma-ray spectra in four energy bins\label{tab:uldiff}} 
\begin{tabular}{ccccccccc}
\hline
$\Delta E$& $N_{\mbox{\tiny{ON}}}/N_{\mbox{\tiny{OFF}}}$ &
$\sigma_{\mbox{\tiny{Li,Ma}}}$ & \multicolumn{6}{c}{$d\Phi^{UL}/dE$ [TeV$^{-1}$cm$^{-2}$s$^{-1}]$} \\
 $[$TeV$]$  & ($N^{UL}_{exc}$) & 95\% C.L. & \multicolumn{6}{c}{($E_*$ [GeV])} \\
& & & $\Gamma=-1.0$ & $\Gamma=-1.5$ & $\Gamma=-1.8$ & $\Gamma=-2.0$ & $\Gamma=-2.2$ & $\Gamma=-2.4$ \\
\hline
$0.1,0.32$ & 51871/52271  & -1.2  & 4.9 $\cdot10^{-11}$ & 5.2$\cdot10^{-11}$ & 5.5$\cdot10^{-11}$ & 5.8$\cdot10^{-11}$ & 6.1$\cdot10^{-11}$ & 6.5$\cdot10^{-11}$ \\
           & (399)        &       & (228) & (211) & (200) & (183) & (187) & (180) \\
  $0.32,1$ &     696/657  &  1.1  & 3.3 $\cdot10^{-12}$ & 3.4$\cdot10^{-12}$ & 3.6$\cdot10^{-12}$ & 3.8$\cdot10^{-12}$ & 4.0$\cdot10^{-12}$ & 4.2$\cdot10^{-12}$ \\
           & (156)        &       & (681) & (631) & (603) & (584) & (566) & (548) \\
   $1,3.2$ &       99/77  &  1.7  & 3.5 $\cdot10^{-13}$ & 3.7$\cdot10^{-13}$ & 3.9$\cdot10^{-13}$ & 4.0$\cdot10^{-13}$ & 4.2$\cdot10^{-13}$ & 4.5$\cdot10^{-13}$ \\      
           & (72)         &       & (2060) & (1917) & (1835) & (1782) & (1730) & (1681) \\
  $3.2,10$ &       69/57  &  1.1  & 5.7 $\cdot10^{-14}$ & 6.0$\cdot10^{-14}$ & 6.4$\cdot10^{-14}$ & 6.6$\cdot10^{-14}$ & 7.0$\cdot10^{-14}$ & 7.4$\cdot10^{-14}$ \\ 
           & (48)         &       & (6750) & (6230) & (5937) & (5751) & (5572) & (5402) \\
\hline
\end{tabular}
}
\end{table}

%%%%%%%%%% FIGURE 3 %%%%%%%%%%%%

%\begin{figure}[t]
%\centering
%\includegraphics[width=0.95\linewidth]{rolkescan.eps}
%\caption{Dependence of the flux UL on the significance of the
%observation for the residual background level of Segue~1 data above 100~GeV.}  
%\label{fig:rolke}
%\end{figure}

% Integral upper limits
% ======================================
\subsection{Integral Upper Limits for power-law spectra}
\label{subsec:intul}
Extending the computation of the ULs to energies above a given
threshold $E_0$, we end up with the following expression:
\begin{equation}
%\Phi^{UL} (>E_{0}) = \frac{N^{UL}_{exc}(E_{rec}>E_0)\;\int_{E_{0}}^{\infty}
%  S(E) dE}
%{t_{eff}\;\int_{0}^{\infty}  A_{eff}(E;E_{rec}>E_{0}) \,S(E) dE}
\Phi^{UL} (>E_{0}) = \frac{N^{UL}_{exc}(E_{rec}>E_0)}{t_{eff}} 
\frac{\int_{E_{0}}^{\infty} S(E) dE}{\int_{0}^{\infty}  A_{eff}(E;E_{rec}>E_{0}) \,S(E) dE}
\label{eq:intupperlimits}
\end{equation}
where $S(E)$ is again the assumed energy spectrum of the source,
$N^{UL}_{exc}(E_{rec}>E_0)$ is derived, as mentioned above, from the
corresponding $|\alpha|-$plot for $E_{rec}>E_0$, and the effective
area is reweighted for the assumed source spectrum.

\begin{figure}[t]
\centering
\includegraphics[width=0.95\linewidth]{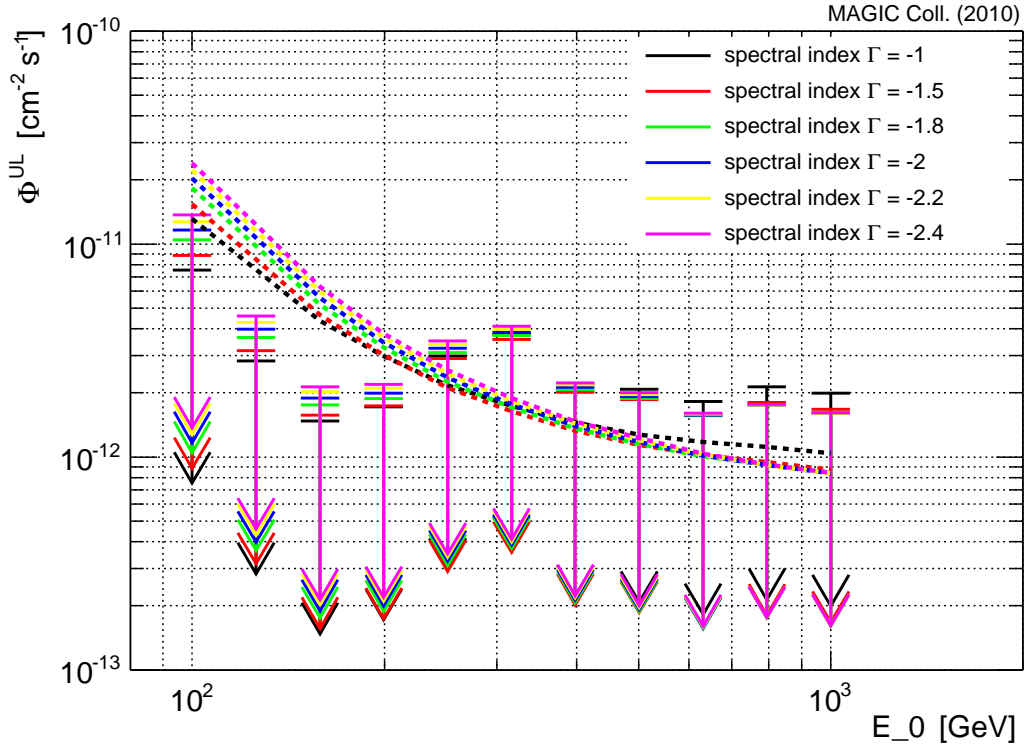}
\caption{Integral flux upper limits from Segue 1. The arrows indicate
  the integral flux upper limits as in
  table~\ref{tab:ulint} for different power-law spectra and energy
  thresholds. On the contrary, the dashed lines indicate the
  corresponding integral upper limits if zero significance
  $\sigma_{\mathrm{Li,Ma}}$ is assumed.}
\label{fig:ulint}
\end{figure}

Table~\ref{tab:ulint} and figure~\ref{fig:ulint} present the integral
ULs achieved by the MAGIC-I observation of Segue~1 for different energy 
thresholds and different power-law spectra with spectral index $\Gamma =
-1.0,-1.5,-1.8,-2.0,-2.2,-2.4$ respectively,  as in the previous section. The results are comparable with
the results of other IACTs on observations of dSphs
\citep{Aharonian:2008a, Albert:2008a, Aliu:2009a, Wood:2008a,
  Aharonian:2009b, Acciari:2010a}.

\begin{table*}[!htb]
\scriptsize{%
\centering
\caption{Integral Segue~1 flux upper limits for several power-law
  spectra and different energy thresholds $E_{0}$ \label{tab:ulint}}
\begin{tabular}{cccccccccc}
\hline
E$_0$ & $N_{\mbox{\tiny{ON}}}/N_{\mbox{\tiny{OFF}}}$ & $N^{UL}_{exc}$ & $\sigma_\mathrm{Li,Ma}$ & \multicolumn{6}{c}{$\Phi^{UL}/\times 10^{-12}$} \\ 
$[$GeV$]$ & & & 95\% C.L. & \multicolumn{6}{c}{$[$cm$^{-2}$s$^{-1}]$} \\
& & & & $\Gamma=-1.0$ & $\Gamma=-1.5$ & $\Gamma=-1.8$ & $\Gamma=-2.0$ & $\Gamma=-2.2$ & $\Gamma=-2.4$ \\
\hline
 100 & 52978/53301 & 453 & -0.99  & 7.5  & 8.8  & 10.5  & 11.6  & 12.7  & 13.7  \\
 126 & 18835/19233 & 174 & -2.04  & 2.8  & 3.2  &  3.6  &  4.0  &  4.3  &  4.6  \\
 158 &   6122/6374 &  93 & -2.25  & 1.5  & 1.6  &  1.8  &  1.9  &  2.0  &  2.1  \\
 200 &   3012/3088 & 110 & -0.97  & 1.7  & 1.7  &  1.9  &  2.0  &  2.1  &  2.2  \\
 251 &   1687/1654 & 194 &  0.57  & 3.0  & 2.9  &  3.1  &  3.2  &  3.4  &  3.5  \\
 316 &   1107/1030 & 250 &  1.67  & 3.8  & 3.6  &  3.7  &  3.9  &  2.9  &  4.1  \\
 398 &     792/761 & 147 &  0.79  & 2.2  & 2.0  &  2.1  &  2.1  &  2.2  &  2.2  \\
 501 &     613/580 & 140 &  0.96  & 2.1  & 1.9  &  1.9  &  1.9  &  2.0  &  2.0  \\
 631 &     536/509 & 124 &  0.84  & 1.8  & 1.6  &  1.6  &  1.6  &  1.6  &  1.6  \\
 794 &     486/445 & 146 &  1.34  & 2.1  & 1.8  &  1.8  &  1.8  &  1.8  &  1.8  \\
1000 &     411/373 & 135 &  1.36  & 2.0  & 1.7  &  1.6  &  1.6  &  1.6  &  1.6  \\
%1585 &     340/320 &  96 &  0.78  & 1.5  &  1.2  &  1.1  &  1.1  &  1.1  &  1.1  \\
\hline
\end{tabular}
}
\end{table*}

%%%%%%%%%% FIGURE 4 %%%%%%%%%%%%

%\newpage
In the case of integral ULs, we see that the dependence on the
different power-law gamma-ray spectra is slightly stronger than in the
differential ULs case, and the UL for the $\Gamma=-1.0$ spectrum is at
most a factor of two stronger than for the $\Gamma=-2.4$ spectrum for
low energy thresholds.

In addition, we clearly see the effect of the statistical fluctuations
(characterized by the significance of detection
$\sigma_{\mbox{\tiny{Li,Ma}}}$) on the integral ULs value. The
significance of detection depends on analysis cuts and the level of
residual background: this has quite a strong impact on the UL of the
number of events $N^{UL}_{exc}$, calculated with the Rolke method, and
eventually on the flux UL (see tables \ref{tab:uldiff} and
\ref{tab:ulint}).  Quantitatively, a smaller significance corresponds
to a lower (more stringent) UL: e.g., going from
$\sigma_{\mbox{\tiny{Li,Ma}}}=-1$ to $1$, the UL gets worse by a
factor of $3$. This is an intrinsic feature of the stastical methods
exploited in the analysis and it should be taken into account when
comparing ULs from different analyses.

For this reason, in order to estimate the effect of the spectral slope and of 
the energy threshold on the value of the integral UL, without being biased
by the fluctuation due to different values of the significance, we computed 
again the ULs assuming a null value for $\sigma_{\mbox{\tiny{Li,Ma}}}$: in
particular dashed lines in figure~\ref{fig:ulint} are obtained with a
significance equal to zero (i.e. with number of ON events equal to the 
number of OFF events in the signal region of the $|\alpha|$-plot) for
different values of $\Gamma$ and $E_0$.

%We finally note that some of the significances calculated here and in the
%next section have negative values. Having checked that no systematics
%effects are present in our data reconstruction, we felt confident to
%use those values. 

%A more conservative approach would have been to set
%to zero all negative significances.
%We also see that for a given gamma-ray spectrum, the integral ULs vary
%considerably for different energy threshold. We will show in the next
%section, how much this can affect the results for typical and real
%gamma-ray signatures from DM annihilation, where the expected spectrum
%is rather different compared to a featureless power-law.  

\section{Constraints on Dark Matter Models} \label{sec:model}
In this section Segue~1 is treated as target for indirect DM searches. 
Assuming a particular form for Segue~1 DM halo, we translate the
ULs calculation described before into constraints on the DM annihilation 
rate. 

The gamma-ray flux due to DM
annihilations depends on $i)$ the intrinsic DM density distribution
in the source, $ii)$ the particle physics characteristics of the DM
candidate  and $iii)$ the telescope energy resolution $\epsilon$, the
field of view $\Delta\Omega$ within which the putative signal is
integrated, and the energy 
threshold $E_0$. It is usually factorized in two terms: 
\begin{equation}\label{eq:flux0}
\Phi(>E_0,\Delta\Omega) = \Phi^{PP}_\epsilon(>E_0) \;
J\,(\Delta\Omega),
\end{equation}
where $\Phi^{PP}_\epsilon$ is the so-called particle physics factor
and reads as follows:
\begin{equation}\label{eq:phipp}
\Phi^{PP}_\epsilon(>E_0) = \frac{1}{4\pi} 
\frac{\langle\sigma_{\mbox{\tiny{ann}}} v\rangle}{2 m^2_\chi}
\int^{m_\chi}_{E_0}\sum^n_{i=1} B^ i \frac{dN^i_\gamma}{dE} \;dE, \\
\end{equation}
where $\langle\sigma_{\mbox{\tiny{ann}}}v\rangle$ is the velocity
averaged annihilation cross 
section, $m_\chi$ is the DM particle
mass, and $\sum^n_{i=1}B^i dN^i_\gamma/dE=dN_\gamma/dE$ is the sum
over all the $n$ possible annihilation channels producing photons ($B^i$ is
the particular branching ratio for channel $i$).

The term $J(\Delta\Omega)$ is the astrophysical
factor and is given by the line-of-sight integral over the DM density
squared within a solid angle $\Delta\Omega$:
\begin{equation}\label{eq:jpsi}
J(\Delta\Omega)=
\int_{\Delta\Omega}\int_{los}\rho^2(r(s,\Omega))\,ds\,d\Omega.
\end{equation}
% 
%Different profiles have been proposed to model dSph DM density
%$\rho(r)$, like the isothermal \citep{Kravtsov:1997a,Persic:1995a},
%the Navarro-Frenk-White \citep{Navarro:1997a} or the Moore
%\citep{Moore:1999a} profiles. among others. 
Motivated by results from cosmological simulations, to describe the
Segue~1 DM density distribution we used the Einasto profile
\citep{Navarro:2010a}: 
\begin{equation}\label{eq:einasto}
\rho_{\mathrm{EIN}}(r)=\rho_s
e^{-2n\;\left[\left(r/r_s\right)^{1/n}-1\right]},
\end{equation}
which produced a good fit to the subhalos simulated by the most recent
$N$-body simulations \citep{Springel:2008b}.
Eq.~(\ref{eq:einasto}) is defined by three parameters: the scale density
$\rho_s$, the scale radius $r_s$ and the index $n$, which
typically ranges from $-3$ to $7$. For the computation of the 
astrophysical factor, we have used
$\rho_s=1.1\times10^{8}$~M$_\odot$kpc$^{-3}$, $r_s=0.15$~kpc and
$n=3.3$,  chosen among the central values after the
marginalization of the likelihood used in Ref.~\citep{Essig:2010a}.
The astrophysical factor is compatible with the recent estimation of Ref.~\citep{SanchezConde:2011ap}.  
%The extent of the Segue~1 stellar
%distribution, and thus a conservative minimum extent of its DM
%halo, is $0.25$~deg \citep{Simon:2010a,Essig:2010a}. 
The astrophysical 
uncertainty on $J(\Delta\Omega)$ has been estimated to be
slightly larger than one order of magnitude at $2\,\sigma$ level
\citep{Essig:2010a}. 

\begin{figure}[t]
\centering
\includegraphics[width=0.99\linewidth]{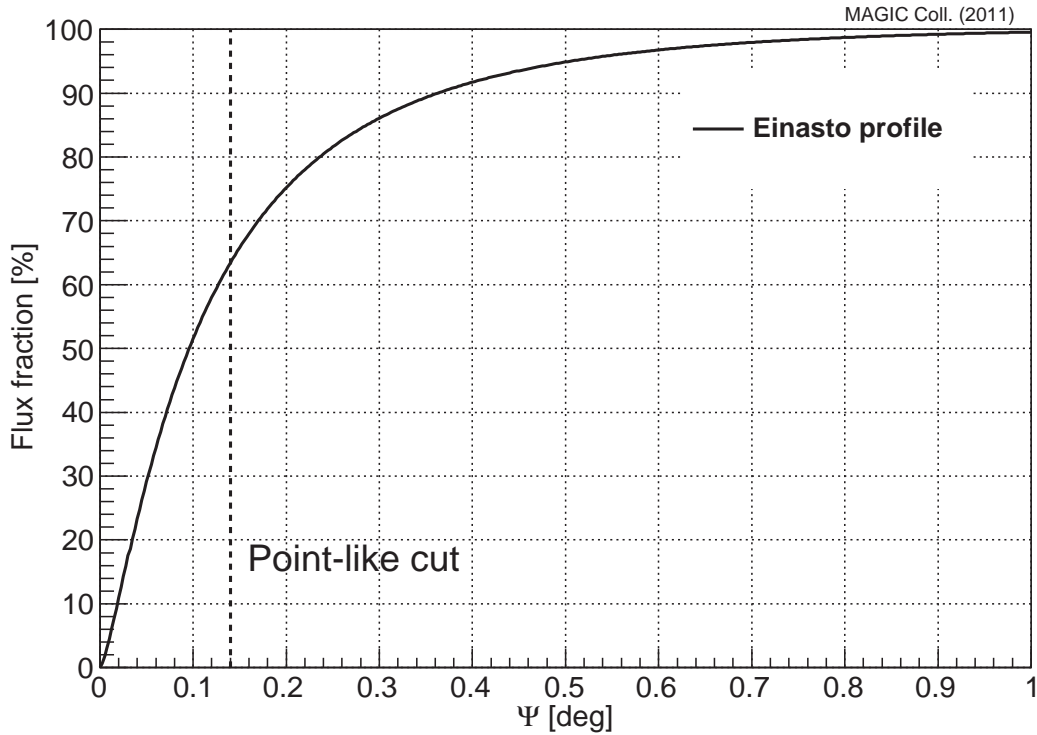}
\caption{\label{fig:perc_graph} Fraction for squared DM density around
  Segue~1 integrated within an angular opening $\Psi$ with respect to
  the total squared DM density, as a function of $\Psi$ itself. The
  dashed line indicates the angular opening corresponding to the
  convolution of the energy dependent $|\alpha|- $cuts used in the
  current analysis, which were optimized for point-like sources. 
  An Einasto profile, as defined in   the text, is assumed to describe
  the DM density distribution.} 
\end{figure}

In figure~\ref{fig:perc_graph} we show how the gamma-ray flux increases
when integrating within larger regions around the center of the
source. We see that the source is slightly extended compared to a
representative MAGIC-I angular resolution of $0.1^\circ$. 
 On the other hand, the contamination of the signal in the
background region, calculated at a distance of $0.8^\circ$ from the
``signal'' region and with the same extension, is well below 1\%. 
% We recall that the background is 
%estimated from a region at a distance of $0.8^\circ$ from the
%``signal'' region and with the same extension. In our case, the signal
%region has chosen to be point-like, because of two reasons. First,
%given the steep falling trend of the astrophysical factor from the
%core of the source, we first expect to see the signal at the center,
%and only in case of strong detection, we will see the source
%extended. Second, we checked that in a circle as extended as the PSF
%of the telescope (point-like hypothesis) where the OFF region is
%estimated, the contribution of the signal to the background is less
%than 1\%.  
The energy
dependent $|\alpha|$ cuts applied in the analysis determine the
angle of integration which fix the fraction of the total astrophysical
factor that must be considered for the ULs computation. In order to
get this angle we have produced a toy MC simulation for
extended sources starting from point-like MC events and spreading the
original gamma-ray arrival directions according to the source DM
density distribution of figure~\ref{fig:perc_graph}. Since the source is
only slightly extended and the assumed luminosity is very peaked at
the center, this method mimics reasonably well a dedicated analysis for an
extended source. We have found that the applied $|\alpha|$ cuts
correspond to an angular integration of $0.14^\circ$ from the Segue 1
center (a solid angle of $10^{-5}$ sr).  From
figure~\ref{fig:perc_graph}, one can see that such angle of integration
encloses $64$\% of the total astrophysical factor.  Therefore, with
the values mentioned above for scale radius and density which leads to
a total astrophysical factor of $J(\Delta\Omega)=1.78\times
10^{19}$~GeV$^2$ cm$^{-5}$ sr, the \emph{effective} astrophysical
factor within the considered analysis cuts is
$\widetilde{J}(\Delta\Omega)=1.14\times10^{19}$ GeV$^2$~cm$^{-5}$~sr.
 This is, to our knowledge, the largest astrophysical factor
among the known dwarf galaxies. We compared in fact our results with
the astrophysical factors of table 4 of Ref.~\cite{Abdo:2010b}, taking care
of integrating over the same region. For a solid angle of
$2.4\times10^{-4}$~sr, we get 95\% of the total astrophysical factor
of Segue~1, i.e. $\widetilde{J}(\Delta\Omega)=1.69\times10^{19}$
GeV$^2$~cm$^{-5}$~sr, while the largest value in Ref.~\cite{Abdo:2010b}
is  $1.2\times10^{19}$ GeV$^2$~cm$^{-5}$~sr in the case of Draco.

%%%%%%%%%% FIGURE 5 %%%%%%%%%%%%

Focusing now on the particle physics factor, we restrict ourselves
for now to the case of a SUSY model in which the presence of a
discrete symmetry ($R-$parity) guarantees that the Lightest
Supersymmetric Particle (LSP) is stable over cosmological timescales
and, therefore, a good WIMP candidate. 
We will consider a 5-dimensional subspace of the Minimal
Supersymmetric Standard Model (MSSM) called mSUGRA
\citep{Chamseddine:1982a, Nilles:1984a}, which is
defined by universal masses for the gauginos ($m_{1/2}$), scalars
($m_0$), and trilinear couplings ($A_0$), as well as by the ratio of the
vacuum expectation values of the two Higgs fields ($\tan\beta$) and
the sign of the Higgsino mass term sign($\mu$).
%
%A generic Minimal SuperSymmetric Model
%(MSSM) requires more than 100 parameters for a complete
%characterization. This number can go down to 5 under certain
%prescriptions: the absence of CP-violating terms and flavour-changing
%neutral-currents, a gravity-mediated SUSY soft-breaking and the
%gaugino mass terms $(M_1,M_2)$ related as $M_1=5\tan\theta_W/3 \sim
%0.5 \mbox{ }M_2$ at the GUT scale.  The 5 remaining parameters that
%define this highly-constrained class of SUSY models called minimal
%SUperGRAvity \citep[mSUGRA,][]{Chamseddine:1982a,Nilles:1984a} are the
%universal masses for gauginos $m_{1/2}$ and scalars $m_0$, the
%universal trilinear coupling $A_0$, the ratio between the vacuum
%expectation values of the two Higgs bosons $\tan\beta$ and the sign
%for the Higgs mass term sign$(\mu)$. 
%
In the majority of mSUGRA models,
the LSP is the lightest neutralino $\chi$, a
linear combination of the super-partners of the gauge bosons
and neutral Higgs bosons.

In order to study the phenomenology of mSUGRA we performed a grid scan
over the parameter space. We spanned (with linear steps) the regions
indicated in table~\ref{tab:scan}, chosen to provide neutralino masses
within the range of detection for MAGIC-I. For each direction we
considered 40 steps and in each bin along $m_0$ and $m_{1/2}$ we
randomly selected a point within that bin.  The scan has been done
once with a positive sign$(\mu)$ and then again with a negative
sign$(\mu)$, for a total of $ 5 \times 10^6 $ points.  For each mSUGRA
model we used \texttt{DarkSUSY 5.0.4} \citep{Gondolo:2004a}
(which includes the virtual internal Bremmstrahlung
effect~\cite{Bringmann:2007nk}) $\;i)$ to 
test if the model is physical, $ii)$ to check if it passes the
Standard Model (SM) experimental constraints implemented in the code
(e.g. LEP bounds on Higgs mass $m_h>114$~GeV, on chargino mass
$m_{\chi^+}>103.5$~GeV and constraints from $b\rightarrow s\gamma$),
and $iii)$ to compute the relic density \citep[with \texttt{Isasugra 7.78}][]{Baer:2003a}.

\begin{table}[!ht]
\centering
\caption{\label{tab:scan} Parameter space for the scan over
  mSUGRA. From Komatsu et al.~\citep{Komatsu:2010a}
  $\Omega_{\mbox{\tiny{DM}}}^{\mbox{\tiny{WMAP}}}h^2=0.1123$ and
  $\sigma_{\mbox{\tiny{WMAP}}}=0.0035$.} 
\begin{tabular}{ccc}
\hline
Parameter & Range & Steps\\
\hline
$ m_0 $       & $50\,; 5000$ GeV  & 40 \\
$ m_{1/2} $   & $0\,;5000$ GeV   & 40 \\
$ \tan\beta $ & $2\,;62$         & 40 \\
$ A_0 $       & $-7000\,;7000$   & 40 \\
sign($\mu$)   & $+;-$               & 2  \\
\hline
\multicolumn{2}{r}{Total number of models scanned} &
  $5.12\times10^6$\\
\hline
\multicolumn{2}{r}{\& passing SM bounds} &  $2.42\times10^6$\\
\multicolumn{2}{r}{\& with $\Omega_{\mbox{\tiny{DM}}}
h^2-\Omega_{\mbox{\tiny{DM}}}^{\mbox{\tiny{WMAP}}}h^2<3\sigma_{\mbox{\tiny{WMAP}}}$} &  $42427$\\
\multicolumn{2}{r}{\& with $|\Omega_{\mbox{\tiny{DM}}} h^2-\Omega_{\mbox{\tiny{DM}}}^{\mbox{\tiny{WMAP}}}h^2|<3\sigma_{\mbox{\tiny{WMAP}}}$} &  $4180$\\
\hline
\end{tabular}
\end{table}

%%%%%%%%%% FIGURE 6 %%%%%%%%%%%%
\begin{figure}[t]
\centering
\includegraphics[width=0.95\linewidth]{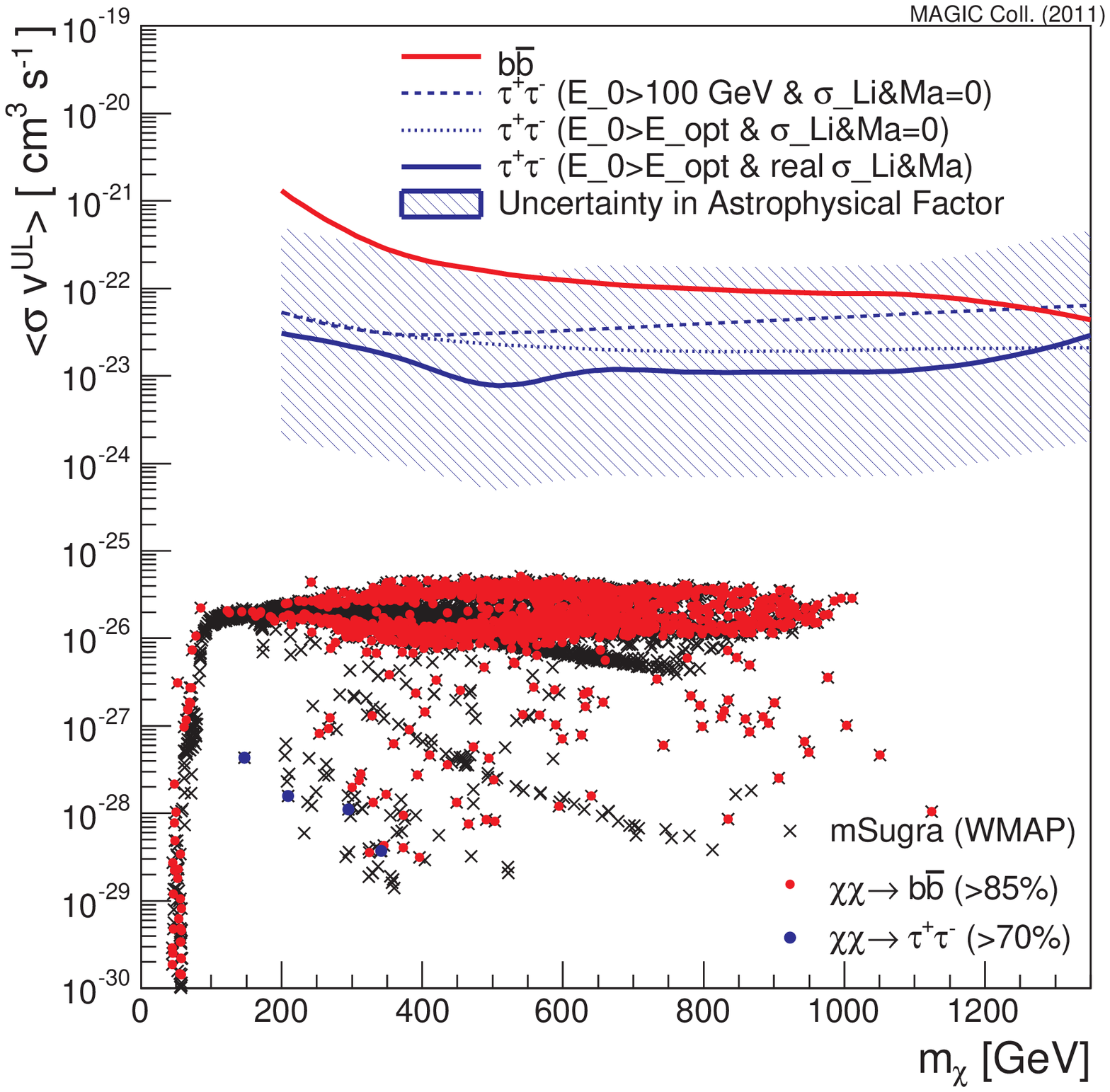}
\caption{Annihilation cross section ULs from Segue 1 MAGIC data
  considering neutralino annihilating entirely into $b\bar{b}$ or
  into $\tau^+\tau^-$. mSUGRA models with a relic density within
  $3\sigma_{\mbox{\tiny{WMAP}}}$ from the WMAP value are plotted
  (black crosses). Among these, neutralinos annihilating mainly in
  $b\bar{b}$ and $\tau^+\tau^-$ are indicated with red points
  and blue points respectively. The solid red line indicates ULs for
  a neutralino annihilating entirely into $b\bar{b}$ while the blue
  lines the case of annihilations into $\tau^+\tau^-$. The dashed blue
  line is calculated for a $E_0=100$~GeV and by imposing a null
  significance. The dotted blue line is also calculated from a
  null significance but above the optimized energy threshold $E_{opt}$
  as explained the text. Finally, the thick solid blue line represents
  the UL calculated with the real significance and above
  $E_{opt}$. The $E_{opt}$ is optimized for all the DM masses. Finally,
  for annihilations into $\tau^+\tau^-$, the blue band covers the
  $2\sigma$ uncertainty on $J_\Theta(\Delta\Omega)$.} \label{fig:sigmav_ul_lines}
\end{figure}

All the models in the scan that correspond to a neutralino with a relic 
density compatible with the value derived by WMAP data within three times its 
experimental error $\sigma_{\mbox{\tiny{WMAP}}}$ \citep{Komatsu:2010a} are 
plotted as crosses in figure~\ref{fig:sigmav_ul_lines}. 
The crosses approximatively 
cover the mass range between 100~GeV and 1~TeV, due to the range that 
we considered for $m_0$ and $m_{1/2}$ in the scan. On the contrary, the 
values for the cross section span many orders of magnitude. They are
mainly concentrated around a value of $10^{-26}$ cm$^3$ s$^{-1}$, but there
are models that, due to particular mechanisms, are characterized
by lower values for the cross section. One example is the prominent
``strip''  that crosses the plane from  
$10^{-26}$ cm$^3$ s$^{-1}$ to around $3 \times 10^{-29}$ cm$^3$ s$^{-1}$ for 
neutralinos that co-annihilate with stops and staus, or the ``tail'' at low 
masses (around 50 GeV).  We stress here that the density of the
points in figures 5 and 6 has no probabilistic meaning and that the
region of points compatible with WMAP may extend until few times $10^{-27}$
cm$^3$ s$^{-1}$ if the range of the scan is extended \citep[e.g. figure 25
of][]{Baltz:2008wd}. 
  
%%%%%%%%%%%%%%%%%% FIGURE 7 %%%%%%%%%%%%%%%%%%%%%%%%%%%%%%%%%%%%%%%%%%
\begin{figure}[t]
\centering
\includegraphics[width=0.95\linewidth]{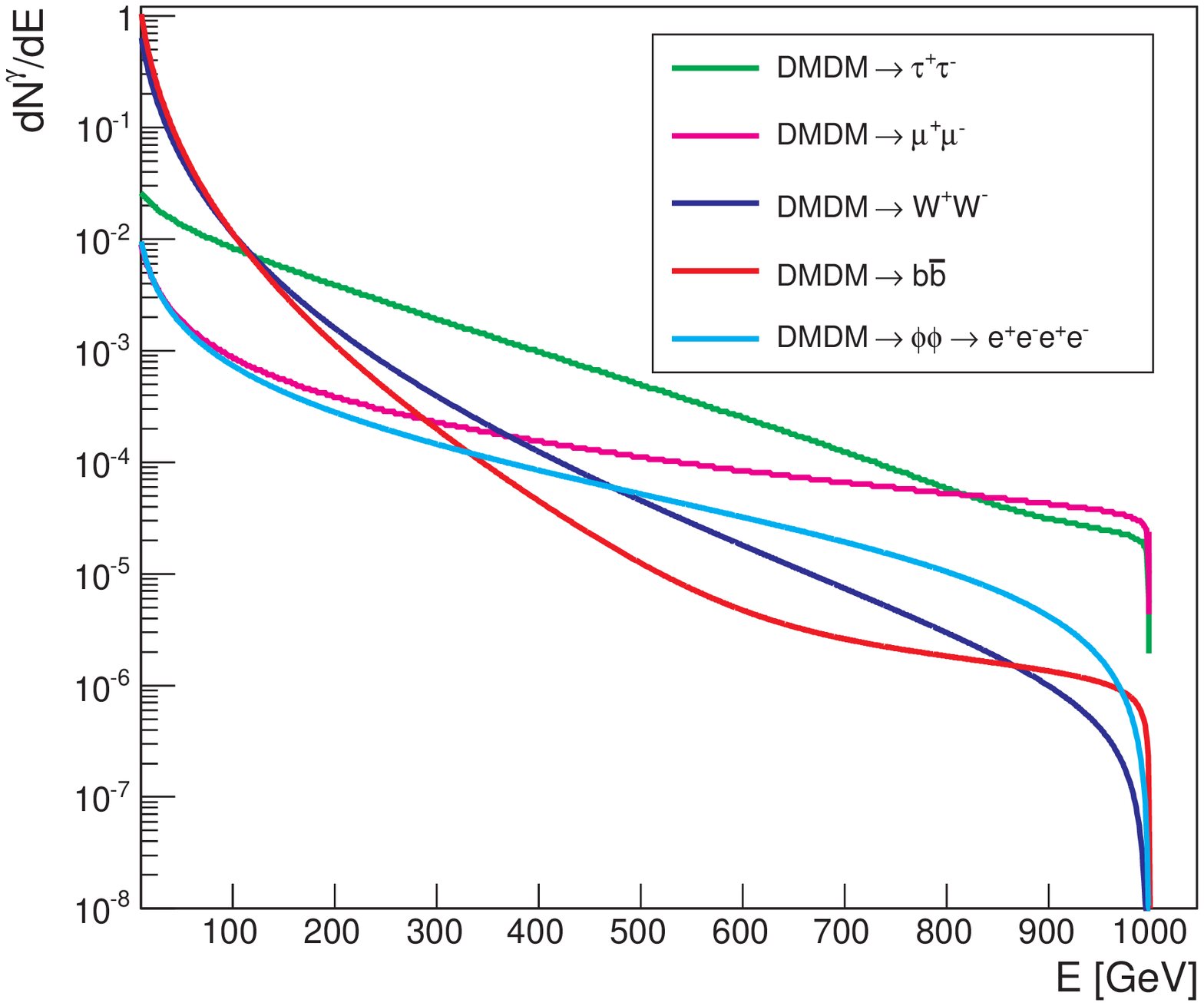}
\caption{Spectra of different annihilation channels. For the
  channels $b\bar{b}, W^+W^-, \tau^+\tau^-, \mu^+\mu^-$ we used the
  fits from Ref.~\citep{Cembranos:2010a}. For $\phi\phi\rightarrow4e$ we
  used the parameterization in Ref.~\citep{Essig:2009a} with $m_\phi=1$~GeV.}
\label{fig:DMspectra}
\end{figure}
%%%%%%%%%%%%%%%%%%%%%%%%%%%%%%%%%%%%%%%%%%%%%%%%%%%%%%%%%%%%%%%%%%%%%%%

%\bred I just moved the below sentence high of one paragraph, the rest
%is the same\ered
For each DM model in the scan, the integral flux UL $\Phi^{UL}(>E_0)$ can be
computed following eq.~(\ref{eq:intupperlimits}), using the Segue~1 data and 
the specific gamma-ray spectrum of the individual DM model.
To ensure a more direct comparison with the particle physics
predictions, we present our results in terms of ULs on the averaged
cross section $\langle\sigma_{\mbox{\tiny{ann}}}v\rangle^{UL}$,
following eq.~(\ref{eq:phipp}): 
\begin{equation}\label{eq:ul_sigmav}
\langle\sigma_{\mbox{\tiny{ann}}}v\rangle^{UL}=
\frac{8\,\pi\,m_\chi^2\,\Phi^{UL}(>E_0)}
{\widetilde{J}(\Delta\Omega)\; 
\int_{E_0}^{m_\chi} \frac{dN_\gamma}{dE} dE}.
\end{equation}

In Sec.~\ref{subsec:intul} we noticed that the integral ULs may change 
as a function of the energy threshold, with lower (more stringent) ULs if
$E_0$ is larger than the experimental one \citep[see also][]{Bringmann:2009a}.
While this variation is  more  predictable in 
the case of power-laws, the situation may be less clear for annihilation 
spectra that contain features and terminate at the DM mass. An example 
of some gamma-ray spectra used in this analysis for a DM mass of 1~TeV 
is shown in figure~\ref{fig:DMspectra}.  
Therefore, for each model we computed the UL for different values of 
energy thresholds $E_0$ among those listed in table~\ref{tab:ulint}. 
 We start by considering the case of null significance
$\sigma_{\mbox{\tiny{Li,Ma}}}=0$, not to be biased by fluctuations in
the real
$\sigma_{\mbox{\tiny{Li,Ma}}}$ defined above. For each annihilation
channel and DM mass, we determine the optimal energy threshold
$E_{opt}$ as the one with the most stringent flux UL among the set of
$E_0$ considered. 

%To ensure an unbiased comparison, we considered the ideal case of null 
%significance $\sigma_{\mbox{\tiny{Li,Ma}}}=0$. For each $E_0$, we used the 
%appropriate average effective area above $E_0$, reweighted with the 
%particular gamma-ray spectrum from DM annihilations. The value of the
%energy threshold producing the lowest (most stringent) UL was then
%assigned to the particular DM model. Finally we computed again for each
%model the UL using the optimal $E_0$ and the analysis cuts described in 
%Sec.~\ref{sec:magic} (thus abandoning the
%$\sigma_{\mbox{\tiny{Li,Ma}}}=0$ hypothesis). 

The effect of the energy threshold optimization can be seen in 
figure~\ref{fig:sigmav_ul_lines}.  In the bottom part of the
plot, we show the models of the scan compatible with WMAP bounds as
black crosses. In addition, two representative subsets are also shown
using a different color coding according to their main annihilation
channel (red points for branching ratio $B(b\,\bar{b})>0.85$, and blue
points for $B(\tau^+\tau^-)>0.7$), which are also representatives of a
soft and hard gamma-ray spectrum respectively (see
figure~\ref{fig:DMspectra}). On the top part of the plot, different
exclusion lines, obtained from the integral ULs, are shown. The blue 
dashed line is the exclusion curve obtained with the integral ULs for the
$\tau^+\tau^-$ annihilation channel, for 
$\sigma_{\mbox{\tiny{Li,Ma}}}=0$ and with $E_0=100$~GeV. The blue
dotted line represents the same curve but now recalculated with the
optimized energy threshold method described above. One can
clearly see that the optimization works at moderate and high DM masses,
because at low DM masses, by increasing the energy threshold one also
loses too many photons. We recall that these two curves do not
represent real data, because are calculated with
$\sigma_{\mbox{\tiny{Li,Ma}}}=0$. The thick blue 
solid line is finally the exclusion curve calculated with the real
significance $\sigma_{\mbox{\tiny{Li,Ma}}}$ and the optimized energy
threshold. One can now see that below roughly 1.2~TeV the curve is
more constraining than the thin blue solid one because in this regime,
the best energy threshold is between 100 and 200~GeV, where the
significance is negative (from table~\ref{tab:ulint}). Above
$1.2$~TeV, the optimized energy threshold is $251$ GeV corresponding to
a positive value of $\sigma_{\mbox{\tiny{Li,Ma}}}$, and therefore the
two mentioned lines cross. Finally, the
red solid line indicates the UL (with optimized energy threshold) in
the case of annihilations only into $b\bar{b}$. In this case, due to
the soft spectrum of this channel, the optimized energy threshold is
almost all the time 100~GeV, apart from the very last points. This is
reasonable, because soft spectra do not gain more by selecting higher energy
photons. 
%
%the two solid blue lines indicate the UL on 
%the annihilation cross section in the case of annihilations only into the 
%$\tau$ channel (i.e. $B(\tau^+\tau^-)=1$). The thin solid blue line represents 
%the calculation of $\langle\sigma_{\mbox{\tiny{ann}}}v\rangle^{UL}$ for a fixed 
%energy threshold of $E_0=100$~GeV, while the thick blue line indicates the
%same UL after the optimization of the $E_0$: it appears that for DM 
%candidates with masses around 500 GeV a factor of 3 in the UL can be gained
%if the {\it optimal} $E_0$ is considered. 
In the same figure, it is also possible to see how the ULs depend on the
shape of the energy spectrum: differences can be
larger than one order of magnitude  \citep[see also][]{Abdo:2010b}
and, as expected, hard spectra are more constraining. 
Finally, in the case of annihilations into $\tau^+\tau^-$, we also indicate
(by means of the blue band) the uncertainty in the astrophysical factor 
at the $2\sigma$ level, showing how strongly it
affects our ULs and prospects for detection. \newline

%%%%%%%%%% FIGURE 8 %%%%%%%%%%%%
\begin{figure}[t]
\centering
\includegraphics[width=0.95\linewidth]{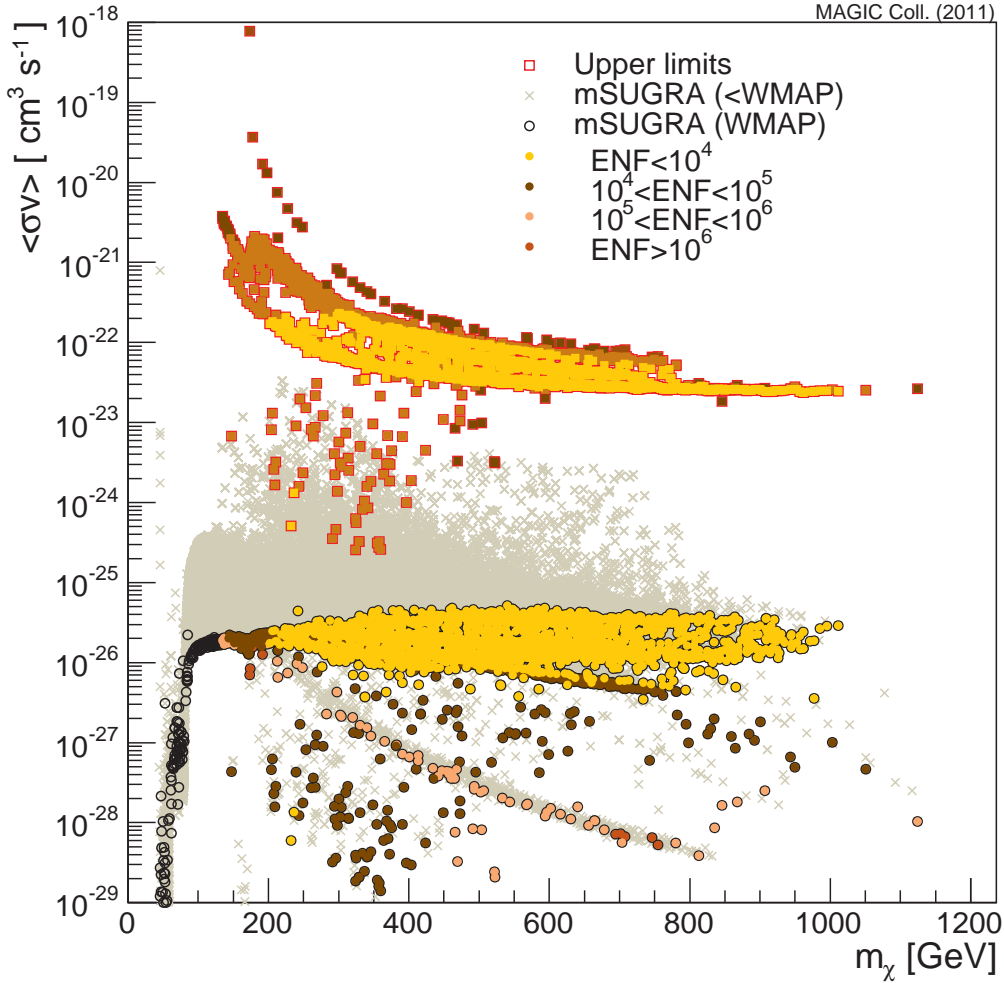}
\caption{Annihilation cross section ULs from Segue~1 MAGIC data
  computed for individual points in the scan. Grey crosses indicate the
  annihilation cross section value for those points in the scan (see
  table \ref{tab:scan}) that pass the SM constraints and with a relic
  density lower than WMAP bound. The full circles only
  consider models within $3\sigma_{\mbox{\tiny{WMAP}}}$ from WMAP. For
  each of these full circles the UL on the cross section can be
  computed from the Segue~1 data (after energy threshold optimization)
  and it is indicated here by a square. Circles and squares are color
  coded in terms of the enhancement factor (see eq.~(\ref{eq:enf})).}
\label{fig:sigmav_ul}
\end{figure}

In figure~\ref{fig:sigmav_ul} we plot again the
$\langle\sigma_{\mbox{\tiny{ann}}}v\rangle$ predictions (same as in
figure~\ref{fig:sigmav_ul_lines}) for the points 
in the scan (full circles), together with the corresponding ULs 
calculated above $E_{opt}$ with the method described above  (full squares). In addition, all the models in the
scan that correspond to a neutralino with a relic density lower than
the value derived by WMAP data plus three times its experimental error
$\sigma_{\mbox{\tiny{WMAP}}}$ \citep{Komatsu:2010a} are also plotted
as grey crosses. In this case, since the standard freeze-out predicts a too small relic
density, we would require a non-thermal production mechanism for the DM
to recover the correct relic density. We are not considering
the case of multi-component DM: if the neutralino is responsible for
only a fraction of the total DM density, it would be reasonable to rescale 
the value of the Segue~1 astrophysical factor (and consequenty the UL) by
the same fraction. 
It can be seen that, for large neutralino masses, ULs
concentrate around $10^{-22}$~cm$^3$~s$^{-1}$, while at lower masses,
the distribution is wider, because models with a low gamma-ray flux
above the threshold are able to produce only loose ULs (as in the case
of squares at low masses with UL of the order of
$10^{-20}-10^{-19}$~cm$^3$~s$^{-1}$).  We stress that each point in the
scan should be compared to {\it its own} UL, and hence the apparent
overlap in figure~\ref{fig:sigmav_ul} between the ULs (plotted as
squares) and some models with a relic density below the WMAP value
(grey crosses) does not imply that any of these models is excluded.
In order to avoid this possible misunderstanding we decide to compute
point per point what we call enhancement factors (ENFs) defined as the
ratio between the UL on the averaged cross section and the value
predicted by mSUGRA:
\begin{equation}\label{eq:enf}
\mathrm{ENF}=\langle\sigma_{ann}v\rangle^{UL}/\langle\sigma_{ann}v\rangle.
\end{equation}
This indicates how much the cross section of the particular model should be
increased in order to make it detectable. In these terms, one can easily
understand which points can be excluded by MAGIC-I data on Segue~1 since 
they would be associated to an ENF smaller than one.
In figure~\ref{fig:sigmav_ul} the color coding is chosen in terms of the 
ENFs: yellow for points with an ENF smaller than $10^4$, orange for models 
with $10^4<$ENF$<10^5$, red if $10^5<$ENF$<10^6$ and brown if ENF$>10^6$.

\begin{figure}[t]
\centering
\includegraphics[width=0.95\linewidth]{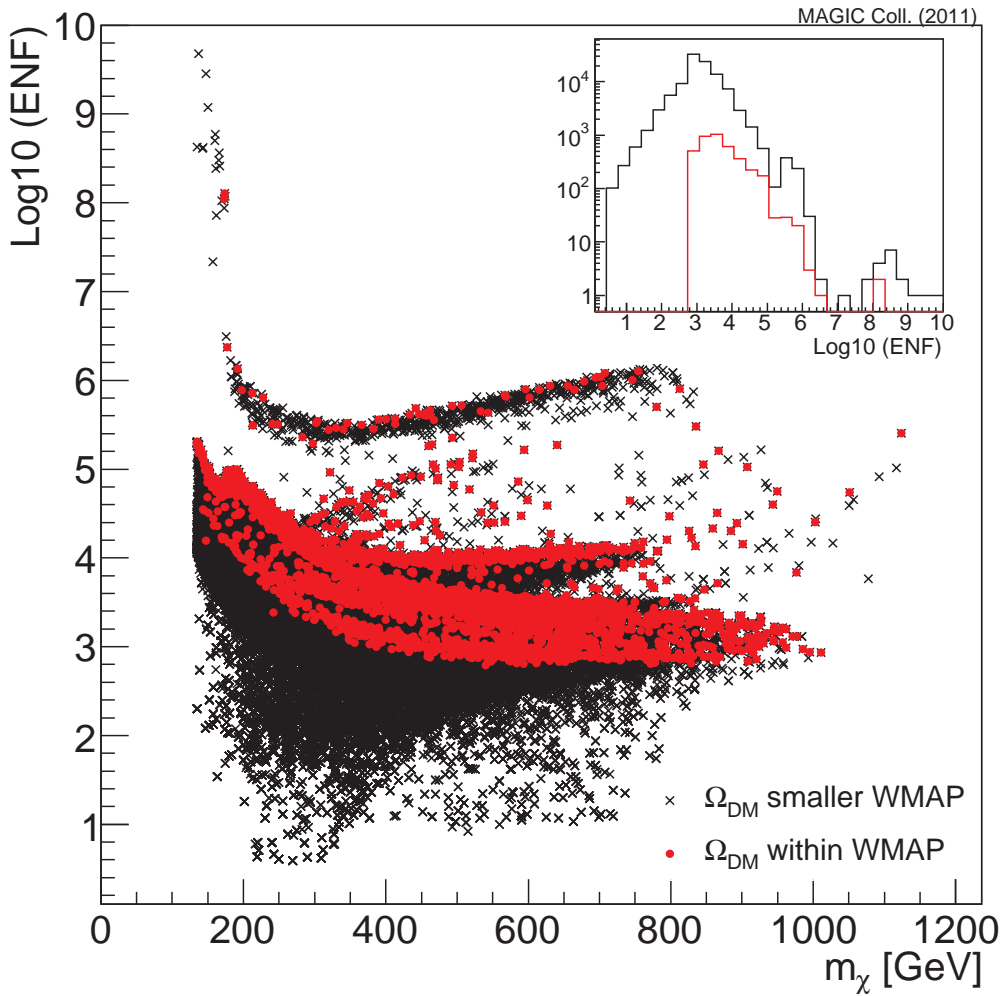}
\caption{Enhancement factors as a function of the DM mass for models in the 
  scan providing a relic density compatible with WMAP value (red crosses) or 
  below (black crosses). The panel in the upper right part indicates the 
  distribution of the ENFs for these two sets of models with the same color
  coding.}
\label{fig:ENFs}
\end{figure}

In figure~\ref{fig:ENFs} the ENFs are also plotted as a function of the mass
for models compatible with WMAP value of the relic density (red crosses)
and below (black crosses). The panel in the upper right of the figure 
indicates the ENF distribution for the two sets of models.
For those compatible with WMAP (red crosses), the lowest ENF is of the order 
of $10^3$. Figure~\ref{fig:sigmav_ul} tells us that the models charactized 
by the lowest ENF are those with the largest annihilation cross section and
a neutralino mass above 200 GeV.
On the contrary, the majority of the points in the scan have an ENF
$>10^4$. Moreover, it can also be seen that the distribution of ENFs
is quite wide. As commented before, this large spread is due to the very 
high (less constraining) ULs relative to models with small neutralino 
masses, as already pointed out in Refs.~\citep{Albert:2008a, Aliu:2009a}, 
and in general to models with a low gamma-ray flux above the energy threshold.
In the case of models with a relic density below the WMAP value (black
crosses), the situation is slightly better: their intrinsic higher
cross sections make them closer to {\it their} ULs, the ENF peaks at
values somewhat lower than those for points compatible with WMAP and
the distribution extends to lower values. 
 We have around $160$ points with ENF$<20$, $40$ with
ENF$<10$ and $5$ with ENF$<5$ even if the exact numbers depend on the
number of points scanned. 

%%%%%%%%%% FIGURE 9 %%%%%%%%%%%%

For the sake of completeness, we point out that the contribution of
monochromatic lines to the annihilation spectrum is neglected in this 
study, due to the fact that their contribution is subdominant with respect
to the continuum emission \citep{Bergstrom:1998a}.

\subsection{Impact on PAMELA preferred region}
\label{sec:pamela}
\begin{figure}[t]
\centering
\includegraphics[width=0.95\linewidth]{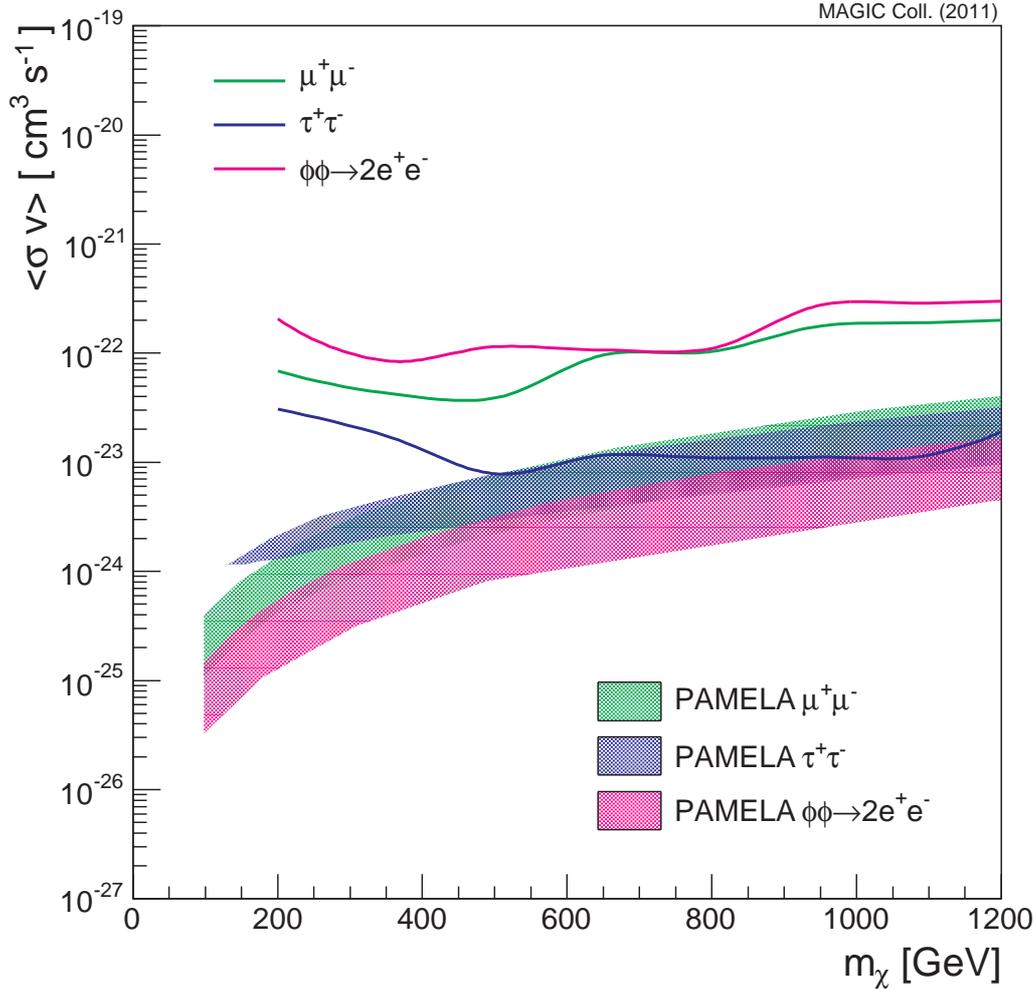} 
\caption{Exclusion lines for a neutralino DM annihilating exclusively into
  $\mu^+\mu^-$ (green lines) or $ \tau^+\tau^-$ (blue line) and for a
  DM candidate interacting with a light intermediate state $\phi$
  decaying into a pair of electrons (pink line). The same annihilation
  channels (with the same color coding) are considered to draw the
  regions in the plane that provide a good fit to the PAMELA
  measurement of the energy spectrum of the positron fraction. The
  regions are taken from Ref.~\citep{Essig:2010a}, which are adapted 
  from Ref.~\citep{Meade:2010a}. We used an astrophysical
  factor of $\widetilde{J}(\Delta\Omega)=1.14\times10^{19}$
  GeV$^2$~cm$^{-5}$~sr.}   
\label{fig:sigmav_ul_pamela}
\end{figure}

Despite the fact that several astrophysical explanations have been
proposed to describe the rise in the energy
spectrum of the positron fraction $e^+/(e^++e^-)$ measured by PAMELA
\citep{Hooper:2008kg, Yuksel:2008rf, Profumo:2008ms, Adriani:2009a}, an interpretation in terms of DM is still
possible \citep[see, e.g.,][]{Cirelli:2008a, ArkaniHamed:2008a,
  Pospelov:2008a, Cholis:2008a, Cholis:2008b, Bergstrom:2009a,
  Meade:2010a, Finkbeiner:2010a}. Typically, the DM particle has to be
heavy and annihilate mainly to leptons. One possibility that has been
largely studied is that the annihilation to leptons occurs through
the production of an intermediate state $\phi$, mediator of a new,
long range, attractive force~\citep{ArkaniHamed:2008a}. However, a
very large annihilation cross section is required, about a factor
$100-1000$ larger than the canonical value derived for thermal
production of $\langle\sigma_{ann}v\rangle \sim
10^{-26}$~cm$^3$~s$^{-1}$.
In the following, we test our ULs against some of the models proposed
in the literature that fit the PAMELA data.
The regions in the $(m_\chi,\langle\sigma_{ann}v\rangle)$ plane that provide 
a good fit to the PAMELA data are shown in figure~\ref{fig:sigmav_ul_pamela} 
for a DM candidate annihilating into $\mu^+\mu^-$, $\tau^+\tau^-$ and for 
the case of the intermediate state $\phi$ decaying to $e^+e^-$, with
$m_\phi=1$~GeV. 
These regions have been adapted from Ref.~\citep{Meade:2010a} after
rescaling from a local DM density of $0.3$~GeV/cm$^3$ to
$0.43$~GeV/cm$^3$ \citep{Salucci:2010qr}. 
%The parametrization of the
%$\phi\rightarrow e^+e^-$ spectrum is the same as in
%\citet{Essig:2010a}.   

%%%%%%%%%% FIGURE 10 %%%%%%%%%%%%

Using again the specific DM annihilation spectra, we plot in
figure~\ref{fig:sigmav_ul_pamela} the ULs 
obtained from the Segue~1 data. We can see that, in this case, the
ENFs needed to meet the PAMELA-favoured region are much smaller than for
mSUGRA,  
and in the case of annihilation into $\tau^+\tau^-$ our ULs are 
probing the relevant regions. However, we recall that the
uncertainty in the astrophysical factor (see
figure~\ref{fig:sigmav_ul_lines}) is large: if future data 
on Segue~1 point to an astrophysical factor close to the upper end
of the currently allowed range, our observations of
Segue~1 might be able to confirm the exclusion of the PAMELA region for
DM particles annihilating in $\tau^+\tau^-$, at least for massive DM 
candidates. Note that for the annihilation into $\phi\phi\to4e$, an
additional Sommerfeld enhancement due to the lower DM velocity
dispersion in Segue~1 compared to the DM velocity dispersion in the MW
halo may effectively shift up the PAMELA preferred region by an order
of magnitude, at least for low enough $\phi$ masses (see also
Ref.~\citep{Essig:2010a}). 

\section{Summary and discussion}
\label{sec:summary}
In the present paper we presented the observation of the
ultra-faint satellite galaxy Segue~1 performed by the MAGIC-I
telescope (single telescope mode) over 29.4 hours of selected data. A
description of the data analysis is reported with the result of no
detection above the background for energies larger than 100~GeV
(sections~\ref{sec:magic}). 

The result is used to compute ULs on the gamma-ray emission from
the source, assuming different power-law energy spectra. The
computation is done first in energy bins (table~\ref{tab:uldiff} and
figure~\ref{fig:uldiff}) and then for integral ULs above different energy 
thresholds (table~\ref{tab:ulint} and figure~\ref{fig:ulint}). In all
cases, we averaged the effective area reweighting it with the specific
gamma-ray spectrum, which allowed us to determine how much the 
ULs depend on the specific spectrum. We also pointed out the fact that
one can get more stringent ULs if computed above energies larger  
than the experimental energy threshold, as a result of the interplay
between the larger sensitivity of the experiment at moderate energies
and the assumed spectrum. 

We focused then on indirect detection of DM and produced
ULs on the annihilation cross section for a large scan of neutralino
models within the mSUGRA scenario (figure~\ref{fig:sigmav_ul}). The ULs
are derived separately for each point in the scan in order to completely account
for the dependence on the specific spectral shape. Results indicate that
ULs are quite dependent on the energy spectrum and a general
exclusion plot cannot be drawn to constrain the parameter space
(figure~\ref{fig:sigmav_ul}). For 
this reason, we find it quite useful to provide the results in terms of
enhancement factors, defined as the intrinsic flux boost needed to meet 
detection (eq.~\ref{eq:enf}). Results are shown in figure~\ref{fig:ENFs}. 
%This plots precisely tells us that current MAGIC-I measurement cannot
%constrain any portion of the parameter space for models with relic
%density compatible with WMAP nor for models with smaller relic
%density. 
A mininum boost is found of the order of $10^3$ (for models compatible with
WMAP) while ``typical'' values are at $10^{4-5}$. However, if we loosen the 
constraint and request only that the SUSY models do not overshoot the 
WMAP value for the relic density, then the situation improves since we can 
have ENFs as low as a few. In these terms MAGIC-I data on Segue~1 are not so 
far from excluding portions of mSUGRA. 

We have also discussed how MAGIC-I data on Segue~1 can be used to test
the PAMELA results on cosmic-rays. 
Current results are probing the PAMELA preferred region for the case
that DM annihilates into $\tau^+\tau^-$. Future improved measurements
of the kinematics of the stars in Segue~1 may decrease the uncertainty
in the line-of-sight integral over the DM density squared and rule out
this region.
%While current results are only
%marginally at tension with one of the models proposed to explain
%PAMELA data (for DM annihilating through the $\tau^+\tau^-$ channel),
%if we consider the most optimistic value for the current astrophysical factor
%within the uncertainties (figure~\ref{fig:sigmav_ul_lines}) and future
%improved Segue~1 kinematical data shift its astrophysical factor
%towards this figure, MAGIC-I data might be able to confirm the
%exclusion of the PAMELA preferred region for this annihilation channel
%almost completely (in the case of massive neutralinos).

The robustness of our results depends mainly on the assumptions on the
astrophysical factor, since an uncertainty of two orders of magnitude
(at $2\sigma$) remains, as estimated in Ref.~\citep{Essig:2010a}.  Our
result can be considered conservative since they do not take into
account intrinsic contributions to the flux from the presence of
substructures in the "smooth" DM halo profile of Segue~1 (which is
already ``per se'' a substructure of the larger MW halo). 
The effects of substructures most likely increase the flux by a factor
of a few at the most \citep{Diemand:2008a, Springel:2008b,
  Martinez:2009jh, SanchezConde:2011ap}. A second contribution 
to the flux, unaccounted for in our calculation, comes from a particle
physics mechanism known as the Sommerfeld effect, which may
additionally boost up the predictions for the gamma-ray emission from
DM annihilation. Its effect on the predicted fluxes from halo
substructures have been studied in Refs.~\citep{Lattanzi:2009a, Essig:2010a}. The most
spectacular effects are present for some resonant values of the DM
mass where the flux can be increased by a factor of $\sim10^4$. How
these models can be constrained by dSphs observation has been studied
in Refs.~\citep{Pieri:2009b, Essig:2010a}. 

Moreover, secondary
gamma-ray emission can be expected, that may 
enhance the predicted flux from Segue~1 and make our ULs more
constraining. As studied in Ref.~\citep{Colafrancesco:2006he}, the main
mechanism for secondary emission is Inverse Compton of electrons and
positrons produced in DM annihilations with CMB and starlight
photons. The contribution of the Inverse Compton with the CMB has been
computed in the case of dSphs in Ref.~\citep{Abdo:2010b} where it can
be seen that the UL on the annihilation cross section can improve of
one order of magnitude for a DM mass around 1 TeV.

It is beyond the goal of this paper to discuss constraints on other DM
scenarios. As a brief comment, we simply recall that 
models with Universal Extra Dimensions predict a stable particle whose 
annihilation signatures are typically harder than the ones in mSUGRA. For 
these models, one can use the ULs we presented for the $\tau^+\tau^-$ 
channel, which can be taken as reference for hard spectra, and therefore 
draw conclusions similar to ours. Furthermore, for the case of decaying 
DM (that can also fit the PAMELA and Fermi/LAT-HESS results), one can rescale 
our results accounting for a different astrophysical factor (depending 
on the DM density and not on the DM density squared) and constrain 
the decay timescale instead of the annihilation cross section.

Comparing MAGIC-I results on Segue~1 with those of Fermi/LAT, the different
energy range covered by the two experiments implies that the latter is
more constraining for low mass DM candidates, while MAGIC (or Cherenkov
telescopes in general) can produce better ULs only for DM heavier than
few hundreds of GeV. For low mass neutralinos, it is already possible for
Fermi/LAT to exclude some of the mSUGRA models with a relic density
smaller than what was measured by WMAP \citep{Abdo:2010b}. With 5 years
of data, Fermi/LAT may be able to probe some of the points 
of the parameter space considered here. In case of no detection, this
will have the effect of excluding most of the models with large ENFs,
shown in figure~\ref{fig:ENFs}, which normally correspond to low-mass
neutralinos. 
 On the other hand, Fermi/LAT will not probe the larger DM masses shown in
figure~\ref{fig:sigmav_ul} and figure~\ref{fig:ENFs}, where IACTs are
more sensitive. In this direction, we recall that with the new
stereoscopic system, MAGIC-stereo allows for improved background rejection
specially below 100~GeV (and therefore higher sensitivity), improved
energy and angular resolutions, and a lower energy threshold
\citep{Colin:2009a}. 

In conclusion, while the observations presented here did not result in a
detection, and the ULs require still quite high (and in some cases 
unmotivated) flux enhancement factors to actually match the experiment 
sensitivity, an analysis like the one presented here is able to point out 
details and features that can be important for future deep exposures of 
this or similar objects, using facilities with much improved sensitivity, 
e.g. the planned Cherenkov Telescope Array \citep{CTAConsortium:2010a}. 

%The first telescope, MAGIC-I, has
%been operating since late 2003, whereas the second one, dubbed MAGIC-II, 
%has started operations at the beginning of 2010~\citep{Cortina:2009a}.

\acknowledgments{
We would like to thank the Instituto de Astrof\'{\i}sica de
Canarias for the excellent working conditions at the
Observatorio del Roque de los Muchachos in La Palma.
The support of the German BMBF and MPG, the Italian INFN, 
the Swiss National Fund SNF, and the Spanish MICINN is 
gratefully acknowledged. This work was also supported by 
the Marie Curie program, by the CPAN CSD2007-00042 and MultiDark
CSD2009-00064 projects of the Spanish Consolider-Ingenio 2010
programme, by grant DO02-353 of the Bulgarian NSF, by grant 127740 of 
the Academy of Finland, by the YIP of the Helmholtz Gemeinschaft, 
by the DFG Cluster of Excellence ``Origin and Structure of the 
Universe'', and by the Polish MNiSzW grant 745/N-HESS-MAGIC/2010/0.\\
The authors wish to thank A.~de la Cruz Dombriz and J.A.R.~Cembranos
for useful discussions and feedbacks  and the anonymous referee for the
careful reading and useful comments on the first version of this
paper. 
}

\bibliography{references}{}
\bibliographystyle{JHEP}

\end{document}